\documentclass{revtex4}
\begin{document}
\title{Geaometry of Quantum Theory: Weyl - K${\ddot a}$hlar Space} 
\author{S. C. Tiwari} 
\address{Institute of Natural Philosophy\\
C/O 1 Kusum Kutir, Mahamanapuri, Varanasi 221 005}
\email{vns_sctiwari@yahoo.com}
\maketitle
%\begin{abstract}
%\end{abstract}
\section{Introduction }
\par The state of a quantum system is represented by a vector in
Hilbert space, and its dynamics can be described by a unitary time 
evolution preserving the inner or scalar product of two vectors in 
the Hilbert space. A notion of distance between quantum states 
and Riemannian structure of Hilbert space are important
geometrical concepts, however until the discovery of Berry p
phase in 1984 physicists generally did not show interest in the
geometry of quantum theory \cite{1}. Experimental observations of Berry 
phase led to important mathematical elucidations of the geometry
of quantum mechanics - in particular the holonomy of fiber
bundles \cite{2}. Physical state space is isomorphic to complex 
projective space which is Kahler, and U(1) part of its holonomy
can be interpreted as the geometric or Berry phase as shown by 
Page [2]. Another geometric effect was discovered in 1990 in the
adiabatic transition probabilty for a two level quantum 
system \cite{3}; a preliminary announcement of Weyl - Kahler space
was made in 1992 \cite{4} as apossible explanation of this effect. The
aim of this paper is to give an exposition of Weyl - Kahler space
and to show that it is a natural geometry of quantum theory.     

Weyl's unified theory aims at a single action principle
to describe electromagnetic and gravitational fields in a 
generalised geometry postulating the principle of Gauge
invariance \cite{5}. In the preface to the first American printing his
book, Weyl writes: "This attmept has failed". Development in quantum
theory led him to reject his original idea of Gauge invariance
though he envisaged "a deep modification of the foundations of 
quantum mechanics". Dirac  in 1973 revived Weyl's theory \cite{6},
however physicists did not show the interest in this. A 
comperhensive discussion on Weyl's theory and a radically new
approach to the meaning of electronic charge are given in a 
recent monograph \cite{7}. Physics does not preclude Weyl's original
gauge transformations since there is a well defined concepts
of the length of a complex vector in the Hilbert space
which may be postulated to acquire length holonomy
under parallel transport. Mathematically it is natural to seek
Weyl structure in complex manifolds. In the next section a 
generalised complex space with hermitian matrix and Weyl gauge 
transformations is discussed. Physically plausible arguments
are presented in Section 3 to suggest Weyl - Kahler space as a
geometrical framework for quantum theory. Prospective application
and open problems constitute the last section.

\section{Weyl - Kahler Space}

Levi - Civita parallelism is a powerful idea in differential
geometry. A vector parallel transported round some closed curve
in a complex manifold gets rotated by some matrix when it returns
to its starting point. The set of matrices for all curves in the
manifold at a given point constitute the holonomy group of the
manifold. For a Kahler space, the holonomy is SU(n) if the 
Riccitensor vanishes (here n is the dimension of the manifiold).
What happens if the length of a vector under parallel transport
also changes ? Weyl space admits such a vectorlength holonomy.
Let us consider a 2n-dimensional real manifold covered by a
system of complex coordinate neighbourhoods such that the transition
functions in the intersection of two neighbourhoods are
holomorphic. Such a manifold is said to admit a complex structure \cite{8}.
We follow Yano's convention \cite{9} and use Latin indices to run over
$1,2,...n, {\bar 1}, {\bar 2},...{\bar n}$ for a complex manifold
$C_n$. A metric in $C_n$ is defined as 

\begin{equation}
ds^2 = g_{ij}dz^idz^j 
\end{equation}  

and a linear ground one form is 

\begin{equation}
 A = A_i dz^i 
\end{equation}

The fundamental metric tensor $g_{ij}$ is assumed to be 
self-conjugate and hybrid, and the covariant vector $A_i$ 
is self conjugate. Weyl gauge transformations are given by

\begin{equation}
 ds \to ds' = \lambda ds 
\end{equation}

\begin{equation}
 A_i \to A'_i = A_i + \partial_i(\ln \lambda)
\end{equation}

Here $\lambda$ is a real scalar function. A complex space
with hermitian metric and gauge field one-form is defined to be a 
Weyl - Hermite space.

A tensor T transforming to  $\lambda^N T$ under the gauge
transformations (3) is defined to be co-tensor of power N: 
$g_{ij}$ is a co-tensor of power 2. The self conjugate mixed
tensor

\begin{equation}
 F_{ij} =
\left[
\matrix{i\delta^\mu_\nu & 0\cr
        0 &  -i\delta^{\bar\mu}_{\bar \nu}\cr}
\right]
\end{equation}
is in-tensor. A skew symmetric self conjugate tensor

\begin{equation}
 F_{ij} = F_i^k g_{kj}   
\end{equation}
is a co-tensor of power 2. In the Weyl-Hermite space generalised
gauge invariant Christoffel symbols $\Lambda_{ij}^i$ can be
derived following Eddington [4], specifically we have

\begin{equation}
^*\Gamma^\chi_{\mu\lambda} = \Gamma^\chi_{\mu\lambda}
 - g^{\chi{\bar \nu}}(g_{\lambda{\bar \nu}}A_\mu +
 g_{\mu{\bar\nu}}A_\lambda)
\end{equation}

\begin{equation}
^*\Gamma^\chi_{\mu{\bar\lambda}} = \Gamma^\chi_{\mu{\bar\lambda}}
 - g^{\chi{\bar \nu}}( g_{\mu{\bar\nu}}A_{\bar\lambda}
 - g_{\mu{\bar\lambda}}A_{\bar \nu}
\end{equation}

\begin{equation}
^*\Gamma^\chi_{\mu{\bar\lambda}} = 0
\end{equation}
and complex conjugates of these. One can calculate curvature
tensors in the usual way. We now define a Weyl - Kahler space
either by deducing a neccessary and sufficient condition such
that $C_n$ and ${\bar C}_n$  are always parallel or postulating
that the covariant derivative of $F_{ij}$ vanishes. Equivalent
conditions are 

\begin{equation}
^*\Gamma^\chi_{\mu{\bar\lambda}} = 0 
\end{equation}

\begin{equation}
\nabla_k F^i_j = 0
\end{equation}

It can be proved that covariant derivative of $F_{ij}$ 
does not vanish, and the two form 

\begin{equation}
F = 2ig_{\mu{\bar\nu}} dz^\mu\Lambda dz {\bar \nu}
\end{equation}
is not closed. Thus unlike Kahler space in which dF = 0, a three
form H is closed in Weyl - Kahler space

\begin{equation}
 dH = 0 
\end{equation}

\begin{equation}
 H = dF = 2A\nabla F 
\end{equation}

A semi Weyl - Kahler space is defined such that 

\begin{equation}
 \nabla_k F^i_j = 0 
\end{equation}

\begin{equation}
 d{\tilde F} = 0 
\end{equation}
where ${\tilde F}$ is a two form $\phi$F; such $\phi$ is a co-scalar
of power -2. An interesting choice is to take scalar curvature
for $\phi$.

\subsection{Remarks}: In mathematical literature conformal class of 
metrics on complex manifolds have been studied by Vaisman and 
Einstein - Weyl structure structure by some groups, see references
cited in [9]. Hermitian - Weyl manifold of Pedersen et al 
is not same as Weyl - Hermite manifold defined here, and seems
similar to Weyl - Kahler. Semi Weyl - Kahler manifold introduced 
for the first time in 1966 \cite{10} has not noticed by mathematicians.
Further our main interest is in the change of length of a vector
under parallel transport.

\section{Quantum State Space}:

A quantum state is a vector in a Hilbert space which for
simplicity is assumed finite dimensional. In Dirac's bra and
Ket notation, let $|\psi(t)>$ be an element of an (N+1) 
dimensional Hilbert space H-{0} i.e. the null vector is 
substracted out. In this complex space one has a scalar product
and a hermitian metric. In component form,

\begin{equation}
 |\psi(t)> = |Z_0, Z_1,...Z_N> 
\end{equation}

Using scalar product, the length of this vector is

\begin{equation}
 <\psi|\psi> = \Delta_{\alpha}{\beta} {\bar Z}^{\alpha}
                 Z^{\beta}  
\end{equation}

A ray is an equivalence class of status for $|\psi> = c|\psi>$ 
where c is a non-zero complex number. The space isometric to complex
projective space $CP^N$. The complex coordinates in $CP^N$ are 
given by 

\begin{equation}
 w^i = \frac{Z^i}{Z^0}  
\end{equation}

For the normalised states $<\psi|\psi>$ is equal to one,
thus the states lies on the unit sphere $S^{2N+1}$. The complex
manifold  $CP^N$ is evolved with the Fubini-Study metric and the
Kahler form (closed) i.e. it is a Kahler manifold. The Kahler
potential is given by 

\begin{equation}
 K = \frac{1}{2}\ln (1+{\bar\omega_i}{\omega^i}) 
\end{equation}

So far we have summarized the standard mathematical 
structure of quantum mechanics. It is non-controversial, and
together with the self-adjoint linear operators representing
physical quantities completes the quantum mechanics. Physical
interpretation of the state vector and quantum measurement 
are the most debated, and as yet unresolved problems. In this
paper we do not intend to address the epistemological and
interpretational questions, however plausible arguments to 
seek a generalisation of the Hilbert space are outlined.
As pointed out in the preceeding section it is mathematically
natural to allow Weyl structure to complex manifold, but
that may not necessarily be useful for physics. There are
atleast two quantum physics phenomena indicating that Weyl - Kahler
space deserves serious consideration as a geometry of quantum
mechanics: 1) geometric adiabatic transition amplitude, and
2) quantum measurements.

Let $|\psi_i>$ and $|\psi_f>$ be two states in the Hilbert
space H, then $<\psi_f|\psi_i>$ is defined to be the transition
probablity amplitude from  $|\psi_i>$ to  $|\psi_f>$. For a 
physical observable the expectation value of the 
corresponding hermitian operator ${\hat S}$  is 

\begin{equation}
 s = \frac{<\psi|{\hat S}|\psi>}{<\psi|\psi>} 
\end{equation}

The expectation value is a real number, and the state vector
in a ray possess the same expectation value. We now examine the
first phenomenon, namely the geometric adiabatic amplitude. For
a two-level quantum systems with a complex hermitian Hamiltonian,
it was shown by Berry that a geometric factor occurs in the transition 
amplitude \cite{3}. Joye et al \cite{11} consider the geometric phase for
a closed path in the complex plane around the eigenvalue 
crossing. The eigenvalues of the Hamiltonian using Teichmuller spaces 
are used to define a metric for the study of the eigenvalues
crossing. Recalling the geometric phase as a holonomy arising
form the parallel transport of a vector around a closed curve,
in $CP^N$, it appears logical to interpret the geometric transition
amplitude as a vectorlength holonomy, see \cite{12} for further discussion
on the theoretical explanation of experiments in optics. 

Theory of measurement is a complex issue; orthodox 
Copenhagers believe that by denying the objective reality to the
wavefunction its collapse or reduction upon measurement becomes an 
empty question. Anandan in the last section  of his paper \cite{13}
argues that one may ascribe ontological or statistical reality
to the wavefunction, and that protective measurement favours
the former as opposed to Einstein's ensemble one. A comperhensive
critique of Copenhogen interpretation by post \cite{14} and the 
most unsatisfactory treatment of single object e.g. an electron
or a photon, in quantum theory \cite{15} show that quantum theory may not
be a fundamental theory. In recent years, new ideas like quantum -
nondemolition measurement and interaction free measurement 
have been proposed; perhaps(!) a variant of many-world-
interpretation (MWI) in the form of consistent histories approach
has also emerged as an alternative to orthodoxy. In the interpretation
due to Omnes \cite{16} the four basic ingredients are consistent histories
decoherence, logic and semi-classical physics. According to him
the last two distinguish his theory from that of others
and branching of the wavefunction of universe a la Everett's
MWI refers to non-issue. Ironically Griffiths does not
find any commonality between MWI and the consistent histories
approach while Gell-Mann identifies 'many-worlds' with
'many-histories' and considers himself to be 'poat-
Everett investigators' see discussion following De Witt's
talk in \cite{17}. Significant contribution in this approach
has been made by Isham \cite{18} towards the mathematical 
structure for a space of generalised histories and decoherence
functionals. According to Griffiths \cite{19} a fundamental
principal of quantum reasonaing may be stated: A meaningful
description of a (closed) quantum mechanical system, including
its time development, must employ a single framework.
For a quantum system a sequence of events is represented
by projectors $E_1$, $E_2$,....$E_n$ on the Hilbert space
at a succession of times $t_1 < t_2.....< t_n$. A history
is represented by a projector 

\begin{equation}
 Y = E_1\otimes E_2\otimes E_n 
\end{equation}
on the history space
\begin{equation}
 {\bar H} = H\otimes H\otimes H\otimes.... H  
\end{equation}
of the tensor product of n copies of H. A consistent Boolean
algebra of history projectors is called a framework. Isham
considers a set of histories and an associated set of 
decoherence functions; this pair replaces the lattice
of propositions and the space of states in orthodox quantum
mechanics. In his framework, an n-time homogeneous history
proposition ($\alpha t_1, \alpha t_2,...\alpha t_n$) can be
associated with a projection operator

\begin{equation}
 Y_1 = \alpha t_1 \alpha t_2...\alpha t_n 
\end{equation}
on the n-fold tensor product of n copies of the Hilbert space
An attractive feature of the consistent histories approach
is that there is no need for a priori time order, and its
potential to address the problem of time in quantum gravity.
Splitting of wavefunction upon measurement and role of time in
quantum mechanics are key factors pointing to the physical 
relevance of Weyl - Kahler space: tensor product space Eq. (23)
can be replaced by a multiple-connected space and topological
index may define a time parameter if quantum-state space is
generalised to Weyl-Kahler.

The novel property of Weyl-Kahler space is obtained 
from Eq.(14). Since H is closed, using Eq. (14) we get

\begin{equation}
 dA = 0  
\end{equation}

Thus A is closed, and the line integral of A round
a loop is zero unless $A_i$ is multivalued. Locally 
$A = df$, therefore nontrivial multiply-connected geometry
results if A is not exact. Assuming topological obstruction
leading to the nonvanishing of the line integral of A, we
have

\begin{equation}
\oint A_i dz^i = n\Lambda
\end{equation}

Here $\nabla$ is some constant, and the coordinates $Z^i$ 
are those given in Eq.(17) i.e. these are the coordinates
on the Hilbert space not the physical space, and the curve on 
this space is parameterised by t. A vector $|\psi>$ parallel
-transported along a curve enclosing the topological 
obstruction (or a branching point) on a Hilbert space H will
jump on to the other connected Hilbert space; let it be denoted 
by  $|\psi>$. We postulate the principal that the length 
change under gauge transformation is the transition probability
$<\phi|\psi>$. Explicitly we have

\begin{equation}
 <\phi|\phi> = \exp(n\nabla)<\psi|\psi>  
\end{equation}

Length holonomy in Eq.(27) for a two-level quantum system
can be interpreted as geometric amplitude factor \cite{3}. In the
problem of measurement, we interpret each successive measuring act
characterized by n, and the exponential factor as a weight 
for multiply connected Hilbert spaces replacing Eq.(23). The
norm of a vector is preserved since each Hilbert space is
assigned a different scale or gauge. Multiply connected Hilbert
spaces have the same dimension, and integer $n$ can be used to define
a time ordering of the measuring events. Since A is closed, a 
natural physical interpretation of the measurement is 'interaction
free process' akin to the Aharonov-Bohm effect. Collapse of
wavefunction is interpreted as a transition of the state
vector to another Hilbert space. A vector which remains on one 
Hilbert space may acquire geometric phase; it is only for
multiply-connected geometry that geometric amplitude and
wavefunction collapse can be accounted for in this approach.

\section{Prospectives and problems}

\par Geometric approach to quantum theory does not imply 
physical reality of wavefunction; in fact there is no evidence
to ascribe wavefunction to a single object without facing
counter-intuitive conclusions or paradoxes. Weyl-Kahler
space for quantum mechanics requires new principles embodied
in Eq.(26) and (27). There is a needto build concrete 
illustrative examples for geometric amplitude and measurement
process since the present paper is limited to conceptual
development of the idea. Tentatively, consistent histories
approach may find the idea of multiply-connected spaces.

Interesting problems from mathematics point of view
also exists; we list some of them. Semi Weyl-Kahler space
defined by Eqs.(15) and (16) is a new geometry, and almost
everything is open for study. The most important probelm
is, of course, the holonomy group classifications of  Weyl-Kahler
manifolds. For non-exact A, Hodge decomposition theoram allows
non-trivial global properties. The nature of topological
obstructions, and examples of multiply-connected Weyl-Kahler
spaces are open problems. In superstring-theories SU(3)
holonomy of Calabi-Yao folds has ben found to be inspired 
by physical arguments; obviously Weyl-Kahler spaces hold
promise for exploring new avenues in this area.

\section{Acknowledgements}
 
I am grateful to M. Atiyah for encouragement, and 
thank to N. J. Hitchin, H. Pedersen and A. K. Pati fro
sending this subject. The library facility at the 
Banaras Hindu University, Varanasi is acknowledged.

\end{document}